\documentclass[letterpaper, 10 pt, conference]{ieeeconf}  

\IEEEoverridecommandlockouts   
\usepackage{amsmath, amssymb, amscd, amsthm, amsfonts}
\usepackage{graphicx}
\usepackage{algorithm}
\usepackage{caption}
\usepackage[font=footnotesize,labelfont=bf,skip=2pt]{caption}
\usepackage[noend]{algpseudocode}
\usepackage{cite}
\usepackage{balance}
\usepackage{bm}
\usepackage{xcolor}
\usepackage{float}
\usepackage{nicefrac}
\usepackage{comment}
\usepackage{booktabs}
\usepackage{colortbl}
\usepackage{xcolor, soul}
\usepackage{stackengine}
\usepackage{paralist}
\usepackage{bbold}

\newtheorem{remark}{Remark}

\newcommand{\id}{\mathbb{1}}
\newcommand{\real}{\mathbb{R}}

\newcommand{\mummsflat}{\mu\text{m}^2/\text{s}}

\definecolor{DarkGray}{rgb}{0.19,0.31,0.31}

\usepackage{stfloats}   

\title{\LARGE \bf
Double-Helix based Real-Time Single Particle Tracking
}

\author{Md Faysal Hossain$^1$ and Sean B. Andersson$^{1,2}$
\\
$^1$Department of Mechanical Engineering, $^2$Division of Systems Engineering, 
\\
Boston University, Boston, MA 02215, USA 
\\
\{faysalh,sanderss\}@bu.edu \\%
}

\begin{document}

\maketitle
\thispagestyle{empty}
\pagestyle{empty}

\begin{abstract}

In Real-Time, Feedback-Driven Single Particle Tracking methods, measurements of the emission intensity from a labeled, nanometer-scale particle are used in a feedback loop to track the motion of the particle as it moves inside its native environment, including within living cells. In this work, we take advantage of Point Spread Function (PSF) engineering techniques that encode the axial position of the particle into the shape of the PSF in the focal plane to eliminate the need for out-of-focal-plane measurements, reducing the complexity of implementation and decreasing the overall measurement time of the control loop. Specifically, we used the Double Helix PSF (DH-PSF) in which a single fluorescent source gives rise to two lobes in the image plane with the lobes rotating in the plane as the particle moves along the optical axis. We designed simple estimators of the relative error between the particle and the tracker, and a simple proportional feedback controller to regulate that error to zero. We explored the efficacy of the approach through simulation studies, demonstrating tracking of fast-moving particles (with diffusion coefficients up to 10~$\mummsflat$) over long time periods (multiple seconds).


\end{abstract}

\section{Introduction}
\label{sec:intro}

Tracking of individual biological macromolecules within the cellular context is a key capability for understanding the dynamics, interactions, and molecular role of nanometer-scale objects, including single proteins, viruses, enzymes, and other sub-cellular components~\cite{liu2020single,scott2023extracting,simon2024guide}. Collectively referred to as Single Particle Tracking (SPT), these methods visualize sub-diffraction-limited sized particles by labeling them with a fluorescent tag, exciting this tag with an excitation source, and then collecting the measured intensity signal. In a sub-class of SPT, known as Real-Time, Feedback-Driven SPT (RT-FD-SPT), the collection of the fluorescence signal is done using a single photon counting module, and a feedback controller actuates the relative position of the detection volume and the sample, essentially locking onto the molecule/particle as it moves. This provides the innate ability to resolve three-dimensional motion as well as higher temporal resolution than in alternative methods based on wide-field camera methods~\cite{van2022real}. Example applications of RT-FD-SPT include recording the interactions of a HIV-1 virus-like particles in living cells~\cite{li2023trajectory}, assessing protein mobility within different regions of the plasma membrane of a cell~\cite{yao2025gradual}, and measuring gene transcription in cancer cells~\cite{donovan2019live}.

\begin{figure}[htbp!]
    \centering\includegraphics[width=0.75\columnwidth]{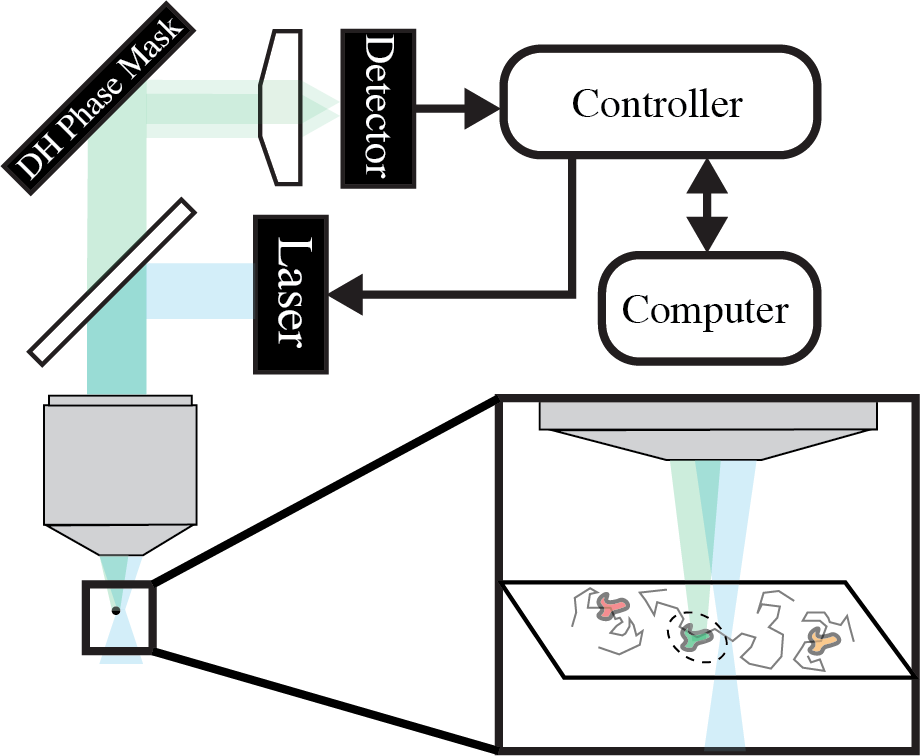}
    \caption{Illustration of the DH-RT-SPT setup.}
    \label{fig:setup}
\end{figure}

The basic structure of all RT-FD-SPT approaches can be understood as a feedback control loop (see Fig.~\ref{fig:setup}) that excites the labeled particle using a focused laser, collects measurements of the emitted intensity from the labeled particle, uses those measurements to estimate the current location of the particle, and then uses a controller to regulate to zero the error between the (estimated) location of the particle and the tracker. Due to the spatial symmetry of the intensity signal in a standard setup, measurements are needed from multiple locations from at least two different planes to be able to determine the location of the particle in all three axes. This can be achieved through placing multiple detectors at multiple locations (as in, e.g., \cite{cang2007guiding}), though the use of multiple detectors significantly increases the cost and the complexity of the optical setup. Alternatively, one can use a single detector and scan the measurement point (as in, e.g.~\cite{ashley2016tracking,hou2020real}). However, these approaches limit overall tracking performance through the necessary assumption that particle motion is slow relative to the scanning speed. Motion along the optical axis is often particularly limiting as it cannot be achieved through beam scanning methods (which can be very fast due to use of acousto- or electro-optic modulators) and is typically either done using piezoelectric actuators (as in \cite{hellriegel2009real}), which are straightforward to implement but relatively slow, the use of multiple excitation beams focused at multiple depths using multiple detectors (as in \cite{mchale2007quantum}), which significantly complicates the experimental setup, increases cost, and increases overall illuminiation of the sample, or tunable acoustic gradient (TAG) lenses (as in \cite{hou2020real}), which are fast but expensive.

There is another class of SPT techniques based on wide-field imaging. These do not leverage closed-loop control but simply collect a series of images for post-experiment analysis (for a discussion of the tradeoffs between wide-field and RT-FD-SPT, see~\cite{van2022real}). Since they are based on camera images, the data is inherently two-dimensional. This challenge led to work on encoding the three-dimensional position of a fluorescent particle into the planar shape of the Point Spread Function (PSF) through PSF engineering~\cite{shechtman2020recent}. One of the early versions was the Double Helix PSF (DH-PSF)~\cite{pavani2009three}. With the DH-PSF,  a single emitter produces s a pair of lobes that rotate about one another as the emitter moves axially. By measuring the angle of rotation of the lobes, the axial position can be determined with nanometer precision, enabling 3D single-particle localization and tracking over an extended depth of field from planar images.

In this paper we utilize the DH-PSF in RT-FD-SPT to achieve three dimensional tracking while only requiring measurements within a single optical plane, reducing the scanning time and thus improving the temporal resolution and tracking rate of an otherwise equivalent system that uses a standard PSF. We label this approach DH-FD-SPT and focus on a simple circular scan pattern combined with linear feedback controllers to achieve particle tracking. The same approach could be used with alternative scan patterns, including the rapid scan in~\cite{hou2020real}.  DH-FD-SPT includes several control hyperparameters that can be tuned to optimize the performance of the tracker with respect to a relevant objective function. Here we focus on tracking time as the ability to track particles for long periods of time is key to understanding their role in cellular processes. 

\begin{figure*}[b!]
  \centering
  \includegraphics[width=\textwidth]{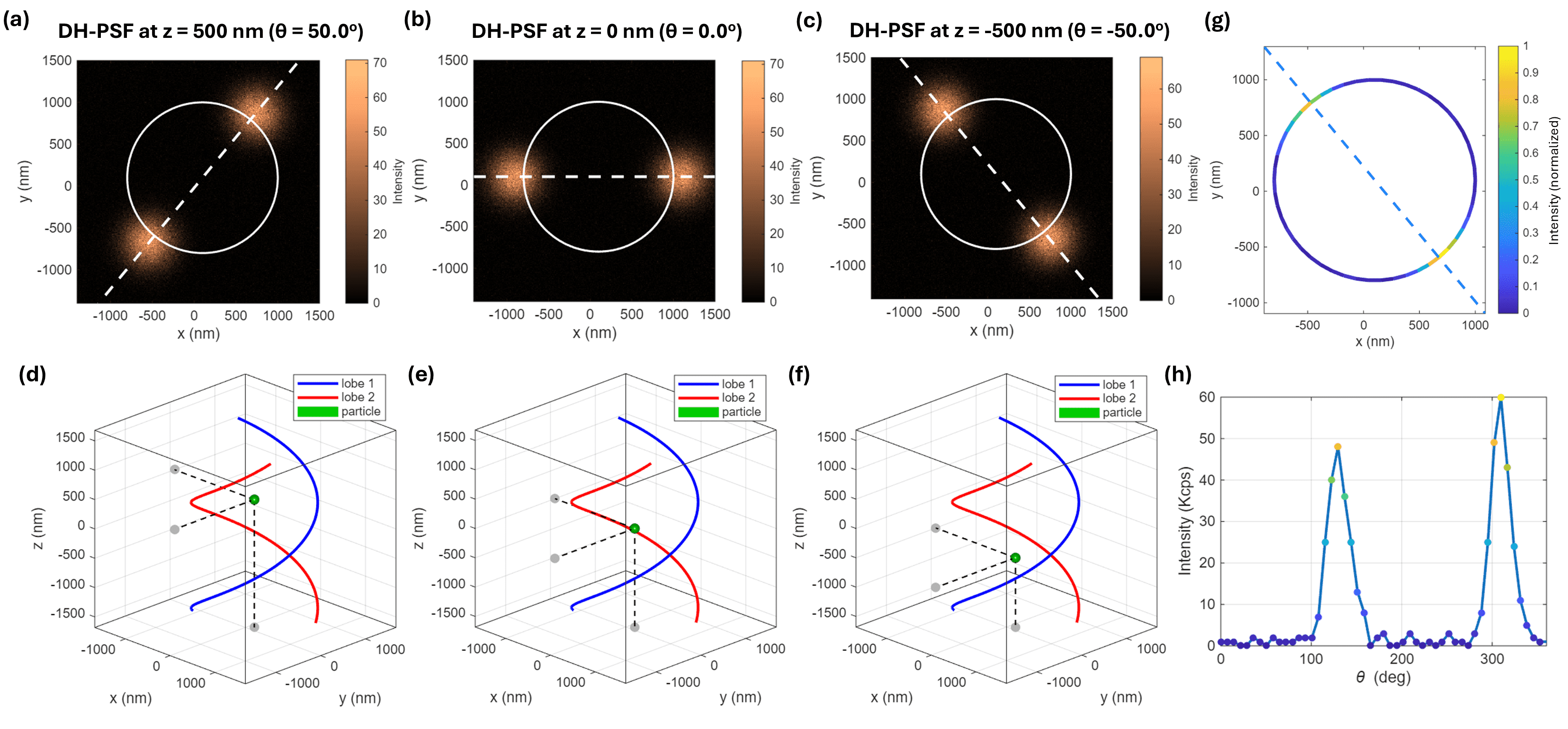}
  \caption{DH-PSF scanning mechanism using parameters $R=1000$ nm, $\sigma_{DH}$ = 200 nm, $r_{scan} = 900$ nm, $G$ = 50, and $N_B = 1.$ (a)-(c) Intensity plots representing the orientation of the DH-PSF along the Z-axis locations of 500 nm, 0 nm, and -500 nm, respectively. The white circle is the scanning radius, and the white dashed lines are the best lines along the double helix.  (d)-(f) shows the particle locations surrounded by the two lobes of the double helix. The tiny green sphere in the middle represents the particle, and the three green dots are the projections of that particle in the xy, yz, and xz planes. The red and blue curves are the trajectories of the two lobes of the DH-PSF. (g) The segmented scanning ring and its intensity variation when z = 0 nm. (h) Intensity variation as a function of angle, $\theta$, at z = -500 nm. }
  \label{fig:mechanism}
\end{figure*}

The main contributions of this work are:
\begin{itemize}
    \item the creation of DH-FD-SPT, a novel algorithm that combines the DH-PSF with RT-FD-SPT to achieve tracking in 3D using scanning in a single plane, simplifying the scanning process and speeding up the controller;
    \item the use of Particle Swarm Optimization (PSO) to optimize control hyperparameters in DH-FD-SPT to achieve long tracking times at low signal intensities;
    \item the demonstration of the efficacy of DH-FD-SPT through numerical studies.
\end{itemize}

\section{DH-RT-SPT}
\label{sec:DH-RT-SPT}

Motion of individual particles within the cellular context can take a variety of forms, including directed motion, anomalous diffusion, and confined diffusion~\cite{godoy2021estimation}. Here, we focus on simple homogeneous Brownian motion, described in discrete time by the model
\begin{equation}
    \label{eq:motion_model_discrete}
    x_p[k+1] = x_p[k] + W[k], \qquad W[k]\sim \mathcal{N}(0,2DT\id), 
\end{equation}
where $x_p = \begin{bmatrix} x_{p_1} & x_{p_2} & x_{p_3}\end{bmatrix}^T \in \real^3$ is the location of the particle, $k$ is the current time step, $W[\cdot]$ is a zero mean white noise process, $\id$ is a 3$\times$3 identity matrix, $D$ is the diffusion coefficient, and $T$ is the discrete time step. Note that we also refer to the position of the particle as $\begin{bmatrix} x_{p} & y_p & z_p \end{bmatrix}^T$ to highlight the axial directions when needed.

The particles are labeled with a fluorescent tag that absorbs the excitation illumination and emits fluorescence photons. The spatial distribution of these photons are described by the PSF of the instrument. In this work we use the DH-PSF which appears as two lobes in the focal plane. Each lobe can be well-modeled using a Gaussian, yielding a spatial intensity $I$ given by
\begin{align}\label{eq:DH-intensity}
    \begin{split}
    I(x,x_p) &= G\exp\left[ \frac{(x_1-x_{1_1})^2 + (x_2-x_{1_2})^2}{2\sigma_{DH}^2}\right] \\
    & +  G\exp\left[ \frac{(x_1-x_{2_1})^2 + (x_2-x_{2_2})^2}{2\sigma_{DH}^2}\right],
    \end{split}
\end{align}
where $x \in \real^2$ is a measurement point in the focal plane, $G$ denotes the maximum signal intensity, $\sigma_{DH}$ is the spread of each lobe, and $x_{1,2}$ are the lobe centers, given by
\begin{align} \label{eq:DH-lobes}
    x_{1,2} = \begin{bmatrix} x_{p_1} \\ x_{p_2} \end{bmatrix} \pm R \begin{bmatrix} \cos \theta \\ \sin \theta \end{bmatrix}.
\end{align}
Here $R$ is a constant defining the distance from the particle position to each lobe center and the angle $\theta$ is determined by the axial position of the particle relative to the focal plane,
\begin{align} \label{eq:DH-theta}
    \theta = -k_{\theta} x_{p_3},
\end{align}
where $k_{\theta}$ is a constant.

The photon generation process is inherently stochastic and as a result the PSF should be interpreted as a rate for a Poisson process. There is also background noise, which we take here to be a constant rate, $N_B$. We assume measurements are made using a confocal modality; in essence, this means a pinhole is used to block out-of-focus light and photons collected using an avalanche photodiode or other single-photon counting module. In practice, this means the number of photons collected at a point $x$ is a Poisson random variable whose rate is the integral of \eqref{eq:DH-intensity} over the area of the confocal detection. For simplicity, here we assume a point measurement so that the measured photons from location $x$ with a particle at location $x_p$ is given by
\begin{align} \label{eq:measuredPhotons}
    I_{\text{meas}}(x,x_p) \sim \textrm{Poiss}\left(I(x,x_p) + N_B\right).
\end{align}
The DH-PSF is illustrated in Fig.~\ref{fig:mechanism}.

Information about the particle location is gathered by collecting photons counts around a scanning circle of radius $r_{scan}$ centered at $x_s \in \real^3$ (see Fig.~\ref{fig:mechanism}). The goal of the tracker is to drive $x_s$ to the particle. Under the assumption that the particle is ``close'' to $x_s$, we calculate an error signal for $x_1$ by taking the difference in photon counts on the left half of the scanning circle from those on the right, and for $x_2$ by taking the difference between the top and the bottom. That is, we define
\begin{align}
    e_{x_1} &= \int_{-\frac{\pi}{2}}^{\frac{\pi}{2}}I_{\text{meas}}\left(x\left(\phi\right),x_p\right)\,d\phi - \int_{\frac{\pi}{2}}^{\frac{3\pi}{2}}I_{\text{meas}}\left(x\left(\phi\right),x_p\right)\,d\phi, \\
    e_{x_2} &= \int_{0}^{\pi}I_{\text{meas}}\left(x\left(\phi\right),x_p\right)\,d\phi - \int_{-\pi}^{0}I_{\text{meas}}\left(x\left(\phi\right),x_p\right)\,d\phi,
\end{align}
where $I_{\text{meas}}(\cdot,x_p)$ is the photon counts defined by $\eqref{eq:measuredPhotons}$ and $x(\phi) = r_{scan}\begin{bmatrix} \cos\phi & \sin \phi \end{bmatrix}^T$ is a point along the scanning circle. (Note that in practice a finite number of measurements is collected and the integral becomes a sum.) An error in the axial direction is determined by estimating the rotation angle $\theta$ by the angle $\hat\theta$ of the line passing through the points along the scanning circle of peak photon counts (subject to the constraint that these two peaks be at least $\tfrac{\pi}{4}$ radians apart around the circle).

The tracking algorithm is then a simple proportional controller,
\begin{align} \label{eq:controller}
    x_s[k+1] = x_s[k] + \begin{bmatrix} k_1 & 0 & 0 \\ 0 & k_2 & 0 \\ 0 & 0 & k_3 \end{bmatrix}\begin{bmatrix} e_{x_1} \\ e_{x_2} \\ \hat\phi \end{bmatrix}.
\end{align}

\begin{remark}
    Intuitively, one may expect better tracking if the orientation of the planar frame given by is aligned to $\hat\theta$ so that motion along $x-$axis is defined to be along the line between the two peaks of the DH-PSF (with the corresponding change to \eqref{eq:controller}. Simulation results, however, indicate that the estimate $\hat\theta$ is noisy and occasionally far off, and that this error can destabilize the controller if it is coupled to the planar directions.
\end{remark}

\begin{figure}[htbp!] \centering\includegraphics[width=1.0\columnwidth]{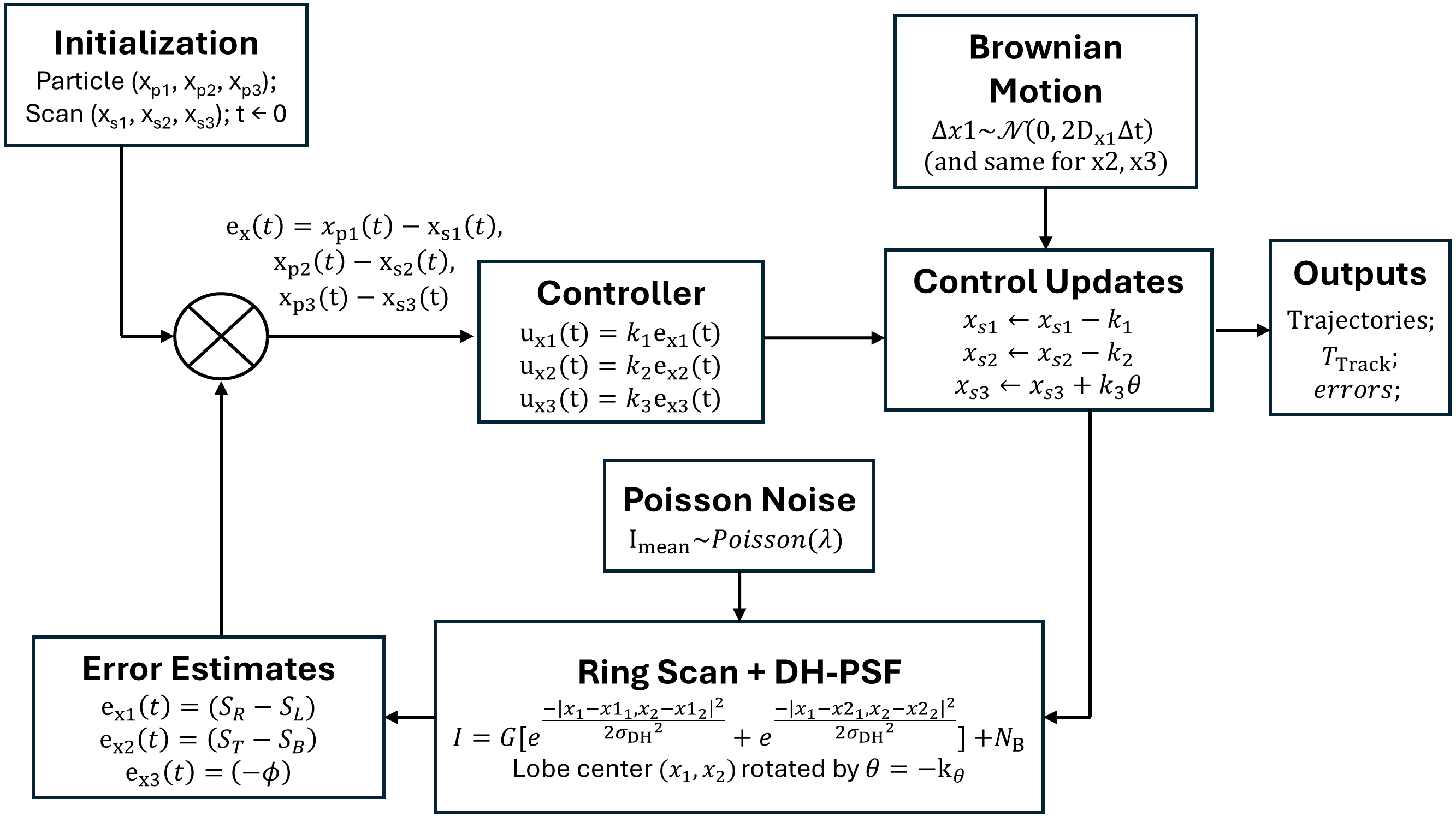} \caption{Feedback control block diagram of a DH-RT-SPT system. } 
    \label{fig:block diagram} 
\end{figure}

The DH-RT-SPT algorithm is summarized in the block diagram in Fig.~\ref{fig:block diagram}.

\section{Parameter optimization}
\label{sec:parameterOpt}

The DH-RT-SPT tracking controller described in Sec.~\ref{sec:DH-RT-SPT} depends on several hyperparameters that can be tuned to maximize performance with respect to specific metrics. The design of the DH-PSF in \eqref{eq:DH-intensity}-\eqref{eq:DH-theta} includes selection of $R$ (the half-distance of the lobe separation), $\sigma_{DH}$ (the spread of the lobes), $k_{\theta}$ (the rotation rate of the lobes as a function of the axial position of the particle), and $G$, the peak intensity. (The background rate is not controllable, though it can be mitigated through good experimental design. Here it is taken to be $N_B = 1$.) The scanning process is defined by $r_{scan}$ (the radius of the scan) and $n_{\text{CirclePts}}$ (the number of points around the circle at which measurements are taken), while the proportional controller is defined by the three gains $k_{1,2,3}$ and the update rate. This large number of parameters makes optimization difficult and so we seek to focus on the ones that are most impactful.

The choice of $k_{\theta}$ is typically dictated by the tradeoff between the desired axial range and the accuracy of determining the $z$ position. Here we fix it at $k_{\theta}=0.1$ rad/nm as a typical value from the literature. Through simulations, we found that increasing $\sigma_{DH}$ generally increased tracking time until about 200 nm, after which further increases led to larger photon counts without any substantive increase in tracking time. We therefore fix $\sigma_{DH} = 200$ nm. The remaining two optical parameters are left to be optimized: the lobe separation (which can be controlled in practice by using a Spatial Light Modulator (SLM) to generate the DH phase mask in real time) and the peak intensity (which can be controlled by adjusting the power of the excitation laser).

Through simulation studies, we found that tracking time was sensitive to $r_{scan}$, $n_{\text{CirclePts}}$, and $k_3$ and thus these were also left to be optimized. Conversely, we found the tracking time was insensitive to the choice of $k_{1,2}$. Based on those results, we selected $k_1 = k_2 = 2.0$. Finally, the controller update is often dictated by hardware bandwidths and compute hardware. Here we fixed the update rate to 2 kHz. The remaining parameters to be optimized over were thus
\[
\Theta = [\, G, \; R, \; n_{\text{CirclePts}}, \; r_{\text{scan}}, \; k_{3} \,].
\]

To find the optimal values for effectively tracking highly diffusive particles, we applied Particle Swarm Optimization (PSO) to maximize the tracking time while minimizing the measured intensity. The motivation for minimizing intensity comes from two facts. The first is that fluroescent labels, particularly organic dyes, photobleach~\cite{zhang2024lumos}. That is, these labels go dark after emitting a finite number of photons. Thus, increasing the photon rate can directly lead to shorter tracking times. The second is that biological systems such as living cells are subject to phototoxicity and limiting its effects are key to obtaining reliable data~\cite{icha2017phototoxicity}.
The specific cost function we used was
\begin{equation}
  J(\Theta) = \mathbb{E}[T_{\text{track}}] 
- \alpha \, \mathbb{E}[\bar{C}]
        - \lambda \, \mathrm{Std}(T_{\text{track}}).
  \label{eq:objective}
\end{equation}
Here, $T_{\text{track}}$ is the achieved tracking duration, defined as the time until the distance error between $x_s$ and $x_p$ exceeded a threshold $\gamma$ (taken here to be 500 nm), $\bar{C}$ is the mean photon counts per scan, Std$(\cdot)$ is the standard deviation, and $\alpha$ and $\lambda$ are user-defined parameters that weight the importance of the photon counts and of the variability of the tracking times.

\begin{algorithm}[tpb!]
\caption{Robust PSO for DH–PSF Tracking }
\label{alg:robust-pso-dhpsf}
\begin{algorithmic}[1]
\Require Weightings $\alpha$ and $\lambda$; error threshold $\gamma$; bounds on parameters $\Theta=\left[G,\, r,\, n_{\text{CirclePts}},\, r_{\text{scan}},\, k_{3}\right]$; required success rate $\rho_\star$; optimization seeds $\mathcal{S}_{\mathrm{opt}}$; validation seeds $\mathcal{S}_{\mathrm{val}}$; maximum restarts $R_{\max}$

\Ensure Optimized parameters $\Theta_\star$ and validation metrics

\Statex \textbf{(A) Particle Swarm Search}
\State Initialize PSO swarm $\{\Theta^{(i)}\}$
\For{iteration $=1:\text{MaxIter}$}
  \ForAll{particles $i=1,\dots,N_{\mathrm{swarm}}$ \textbf{in parallel}}
     \State $\text{score}^{(i)} = J\left(\Theta^{(i)}\right)$
     \label{line:robust-obj-call}
  \EndFor
  \State Update particle velocities/positions; Project to $\Theta^{(i)}$
\EndFor
\State $\Theta_\star \gets$ best particle (max score); refine by \texttt{fmincon}

\Statex \textbf{(B) Validation on Fresh Seeds}
\State Evaluate $\Theta_\star$ on fresh seeds $\mathcal{S}_{\mathrm{val}}$ → ($\rho$, $\mathrm{mean}(T_{\text{track}})$ , $\mathrm{mean}(\bar{C})$, $\mathrm{MAE}_{x,y,z}$).
\State If $\rho < \rho^\ast$, restart PSO (steps 1–6) up to $R_{\max}$ times and revalidate.

\Statex \textbf{(C) Outputs}
\State \Return $x_\star$ and $(\rho, \mathrm{mean}(T_{\text{track}}) , \mathrm{mean}(\bar{C}), \mathrm{MAE}_{x,y,z})$
\end{algorithmic}
\end{algorithm}

Based on prior experience, performance of RT-FD-SPT algorithms in the literature, and simulation results, tracking of particles diffusing at diffusion coefficients $D$ below 0.1~$\mummsflat$ is relatively easy and is insensitive to parameter choice. We thus focused on $D \geq 1.0\,\mummsflat$ for the optimization. In addition, parametric bounds were imposed on the optimization vector $\Theta$ to ensure physically realizable and numerically stable PSO results. On the basis of the simulation results, the decision vector was constrained as follows: $G \in [5,100], \; r \in [700,1200], \; n_{\text{CirclePts}} \in [15,70], \; r_{\text{scan}} \in [700,1200], \; k_z \in [10,220]$.

To ensure robustness given the stochastic nature of particle motion, each set of candidate solutions was evaluated over 20 independent trajectories, defined by a set of seeds,  $\mathcal{S}_{\mathrm{opt}}$, for the pseudorandom variables. 
After simulating the DH-PSF algorithm for each trajectory, the tracking time, photon counts, and success/failure outcomes was recorded and $J(x)$ evaluated based on the aggregated results. In the PSO process, we used a swarm size of 64, terminating optimization when either the particle velocities collapsed to (near) zero, indicating the swarm had converged, or when 150 iterations was reached. After PSO converged, a hybrid local refinement was applied using the Matlab function $fmincon$. In this step, the Monte-Carlo seeds defining the pseudorandom numbers in the simulation were held fixed to further refine the best candidate solution. 


The final optimized solution candidate, was then validated against a new set of $N_s$ trajectories, defined by a set of seeds $\mathcal{S}_{\mathrm{val}}$, and the success rate $\rho$ calculated according to 
\[
\rho = \frac{1}{N_s} \sum_{i=1}^{N_s} \mathbb{1}\!\left[T_s^{(i)} > t_{val}\right].
\]
For diffusion coefficients up to 7 $\mummsflat$, the validation time was set to $t_{val} = 9$ sec. Above 7 $\mummsflat$, the validation time was reduced to 1 second. If $\rho$ was greater than a given threshold $\rho^*$ (taken here to be 0.9), optimization was declared completed. If it was less than this threshold, optimization was restarted. This was repeated a maximum of $R_{\max}$ times (taken here to be 3) at which point the parameters giving the highest success rate were taken as the best values. This optimization approach is summarized in Algorithm~\ref{alg:robust-pso-dhpsf}.

\section{Numerical results}
\label{sec:numerical}

To evaluate performance of the optimized controllers, we consider not only the tracking time achieved but also the average intensity, measured in Kilocounts per second (Kcps), and the tracking error both in the focal plane ($E_{xy}$) and in all three dimensions ($E_{xyz}$),
\begin{subequations}\begin{align}
    E_{xy} &= \sqrt{(x_s-x_p)^TM(x_s-x_p)}, \\
    E_{xyz} &= \sqrt{(x_s-x_p)^T(x_s-x_p)},
\end{align}\end{subequations}
where $M$ is a diagonal matrix with diagonal entries $\begin{pmatrix} 1, 1, 0\end{pmatrix}$. Note that while parametric optimization was performed for a maximum tracking duration of 9 seconds, the maximum simulation time for evaluation was set to 30 sec. 

\begin{figure*}[htbp!]
  \centering
  \includegraphics[width=\textwidth]{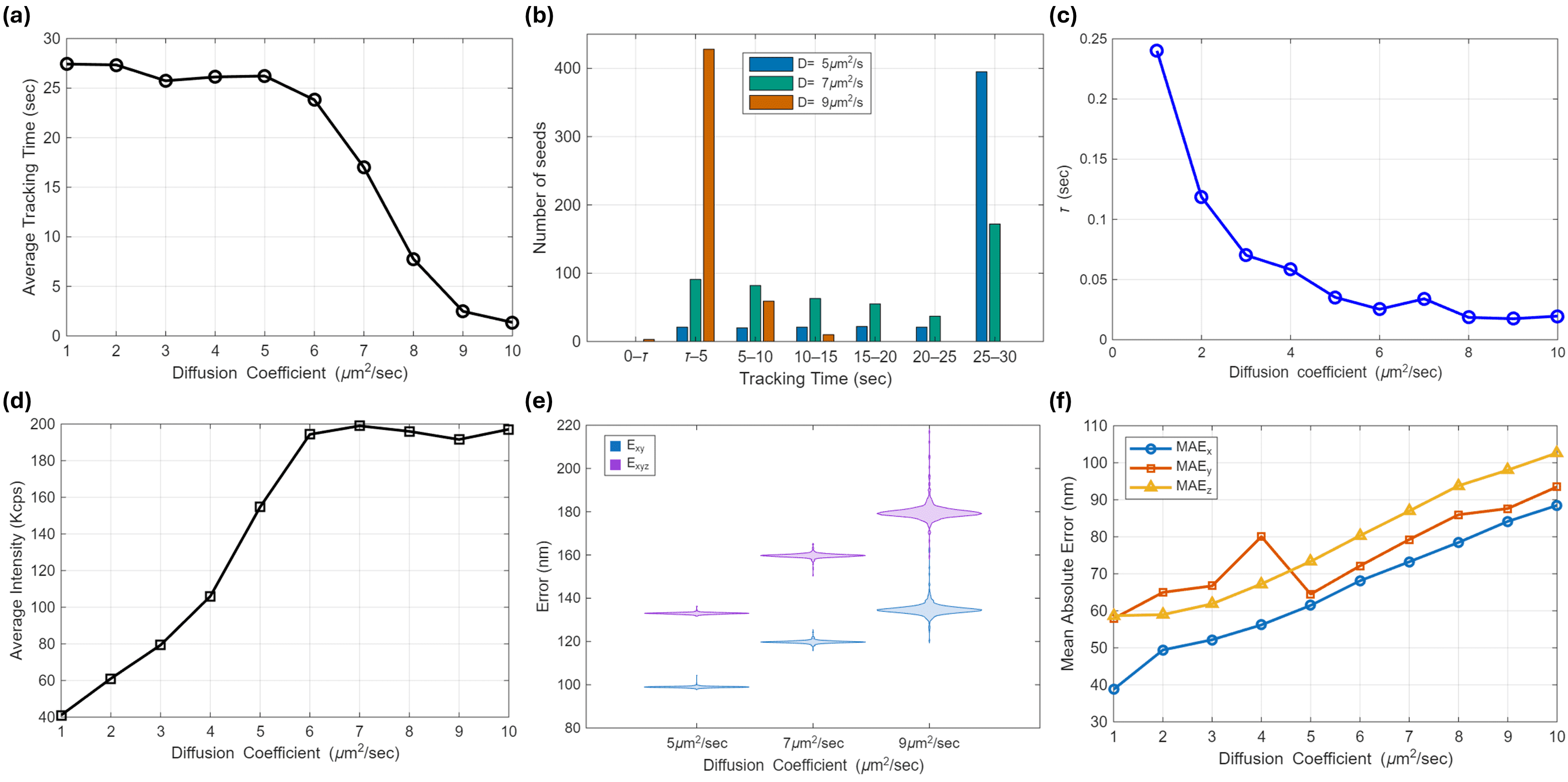}
  \caption{Performance of DH-PSF-based tracking microscope evaluated from 500 independent simulations. (a) Average tracking duration for different diffusion coefficients (1- 10~$\mummsflat$). (b) Histogram of tracking times for diffusion coefficients of 5, 7, and 9 $\mummsflat$. $\tau$ is the first passage time of a diffusing particle out of the scan circle. The 0-$\tau$ column shows the runs that do not result in successful tracking. (c) First passage time ($\tau$) out of sensing area as a function of diffusion coefficient. (d) Average intensity as counts per scan for different diffusion coefficients. (e) Average mean lateral (blue) and combined 3D tracking error (purple). (f) Mean absolute error (MAE) in the individual axes over the change of diffusion coefficients. }
  \label{fig:performance}
\end{figure*}

To ensure the tracker is actually achieving a tracking lock, tracking time should be compared against a baseline of the expected first passage time $\tau$ of a particle starting at the center of the scanning circle to diffuse out of the sensing area, given by 
\begin{equation}\label{eq:tau}
\tau = \frac{R^{2}}{6D},
\end{equation}
where $R$ is the half-distance between the DH-PSF lobes. Note that since $R$ is at optimization variable, its value depends on the diffusion coefficient; the corresponding values of $\tau$ as a function of $D$ are shown in Fig.~\ref{fig:performance}(c). Note that all the first passage times are less than 0.25 seconds.

\begin{figure*}[htbp!]
  \centering
  \includegraphics[width=\textwidth]{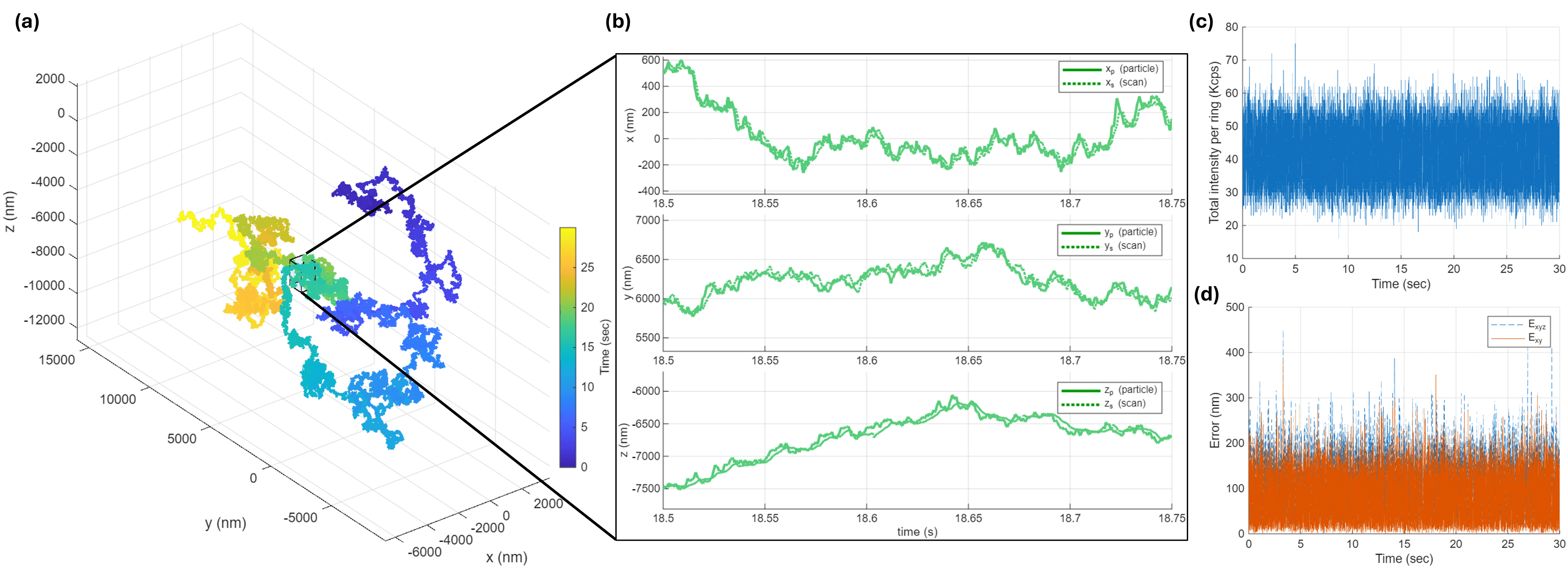}
  \caption{ DH-FD-SPT based microscopy technique.  (a) 3D trajectory of a particle diffusing at D = 1~$\mummsflat$ tracked for 30 sec. The color coding of the curve represents elapsed time. (b) Movement of the particle and the scanning center of DH-PSF in the time frame of 18.5-18.75 seconds. (c) The total measured intensity around each circular scan and (d) the lateral and axial error over time.}
  \label{fig:4D-SPT}
\end{figure*}

A typical example run for a particle diffusing with $D=1~\mummsflat$ is shown in Fig.~\ref{fig:4D-SPT}. The optimized parameters used for this run were $G$ = 22.565, $R$ = 1200 nm, $n_{\text{CirclePts}}$ = 15, $r_{\text{scan}}$ = 971.9 nm, and $k_{z}$ = 62.19. In this run, the tracker followed the particle for the full 30 seconds with stable photon counts per lobe averaging $\approx$ 40 Kcps. Average errors were $E_{xy} = 76.5$ nm and $E_{xyz} = 104$ nm.

We evaluated the performance of the optimized DH-RT-SPT over 500 random trajectories (using pseudorandom seeds distinct from those used in the optimization) with diffusion coefficients $D \in [1,10]$~$\mummsflat$ in steps of 1 $\mummsflat$. Simulations were run until either the tracking error was larger than 500 nm or the tracking time reached 30 s. The results are shown in Fig.~\ref{fig:performance}. The mean tracking times in Fig.~\ref{fig:performance}(a) was nearly 26 sec up until $D = 5$~$\mummsflat$ before falling off. Even at $D=10$~$\mummsflat$, the average tracking time of 1.36 seconds is approximately sevenfold greater than the first passage time. The tracking time histograms in Fig.~\ref{fig:performance}(b) indicate that at $D=5$~$\mummsflat$ the tracker almost always tracked the particle for the maximum duration (with 395 of 500 runs falling into the maximal bin). The distributions then skew towards shorter tracking times as the diffusion coefficient is increased, with no runs lasting more than 15 seconds at $D=9$~$\mummsflat$, though nearly all of them still significantly exceeded the first passage time.

Fig.~\ref{fig:performance}(d) shows the average intensity needed to track a diffusing particle. The results show an essentially linear increase in optimal average intensity until about 6~$\mummsflat$, after which increasing the count rate did not lead to better performance. In general, increasing the count rate leads to an increase in signal-to-noise ratio, likely leading to better error estimates and thus better controller performance. These results indicate, however, that this effect diminishes at high diffusion rates. Beyond 6~$\mummsflat$, raising the intensity did not improve tracking time, leading to a stabilized average counts and a falling mean tracking duration.    


The tracking errors, shown in Fig.~\ref{fig:performance}(e), and the mean absolute error in the three axes, shown in Fig.~\ref{fig:performance}(f), both show that, as expected, tracking becomes more challenging for higher diffusion coefficient. As $D$ increases, both the average displacement of the particle and the variability of that displacement increase, leading to increased error in tracking and, eventually, loss of tracking. The results also indicate that tracking error in the axial direction is always larger than in the plane. This is driven primarily by the fact that estimation of the angle of the lobes, and thus of the $z$ position of the particle, is less robust than the comparison of counts between two halves of the scanning circle.       

\section{Conclusion}
\label{sec:conclusion}

In this paper we introduced a new RT-FD-SPT algorithm, DH-RT-SPT, based on the double helix point spread function. This approach allows us to collect output intensity measurements from only a single plane while still achieving tracking in three dimensions. This simplifies the controller and reduces the time it takes to produce an estimate of the tracking error, potentially leading to an increase in tracking speed over the current state-of-the-art. Controller parameters were optimized using PSO, and performance in terms of tracking time, tracking error, and photon counts was evaluated through simulation. Future work includes implementing the approach on hardware and exploring the efficacy of DH-RT-SPT through a physical experiment.

\section*{Acknowledgements}

This work was funded in part by the NIH through 1R21GM157695-01. 

\bibliographystyle{IEEEtran}
\balance
\bibliography{Andersson_bibs}

\end{document}